# Picosecond Ultrasonics with Miniaturized Semiconductor Lasers

Michal Kobecki,[1] Giuseppe Tandoi,[2] Eugenio Di Gaetano,[2] Marc Sorel,[2] Alexey V. Scherbakov,[1,3] Thomas Czerniuk,[1] Christian Schneider,[4] Martin Kamp,[4] Sven Höfling,[4] Andrey V. Akimov,[5,*] Manfred Bayer[1,3]

[1] *Experimentelle Physik 2, Technische Universität Dortmund, D-44221 Dortmund, Germany*

[2] *School of Engineering, University of Glasgow, Glasgow, G12 8QQ, United Kingdom*

[3] *Ioffe Institute, 194021 St. Petersburg, Russia*

[4] *Technische Physik, Universität Würzburg, 97074 Würzburg, Germany*

[5] *School of Physics and Astronomy, University of Nottingham, Nottingham NG7 2RD, United Kingdom*

**Abstract:** There is a great desire to extend ultrasonic techniques to the imaging and characterization of nanoobjects. This can be achieved by picosecond ultrasonics, where by using ultrafast lasers it is possible to generate and detect acoustic waves with frequencies up to terahertz and wavelengths down to nanometers. In our work we present a picosecond ultrasonics setup based on miniaturized mode-locked semiconductor lasers, whose performance allows us to obtain the necessary power, pulse duration and repetition rate. Using such a laser, we measure the ultrasonic echo signal with picosecond resolution in a Al film deposited on a semiconductor substrate. We show that the obtained signal is as good as the signal obtained with a standard bulky mode-locked Ti-Sa laser. The experiments pave the way for designing integrated portable picosecond ultrasonic setups on the basis of miniaturized semiconductor lasers.

1. Introduction

Picosecond ultrasonics (PU) is an advanced research field which aims to extend traditional acoustic techniques to the gigahertz (GHz) and terahertz (THz) frequency ranges [1,2]. It started in the late 1980ies with the development of ultrafast lasers and nowadays PU shows great potential for elastic nanoscopy [3] and ultrafast control of electronic and optical devices [4-6]. PU imaging with sub-nanometer depth resolution is used to study nanostructures including biological cells [7-10], chemical reactions [11], adhesion of nanolayers [12,13] and profile inhomogeneity [14].

The main element of a PU setup is a pico- or femtosecond laser which output is split in two beams used for pumping and probing the system of interest. In most of PU experiments the pump beam is focused onto a thin metal film which acts as an optoelastic transducer. After being hit by the pump pulse, the film expands in a time of ~1 ps and, as a consequence, a coherent elastic wave packet with frequencies up to several hundreds of GHz is injected into the specimen or device. The critical parameters for PU is the energy density $J$ in the focus spot of the laser pump pulse, the pulse duration $\Delta t$, which limits the temporal resolution and correspondingly the maximum frequency in the ultrasonic wave packet, and the laser repetition rate $f$, which determines the accumulation time. The values for $J$, $\Delta t$ and $f$ that can reliably be reached, cover a wide range ($J=10^{-6}$-$10^{-2}$ J/cm$^2$, $\Delta t \leq 1$ ps and $f=10$-$10^9$ Hz) using different experimental schemes and can easily be achieved using commercial femtosecond lasers. However, all PU setups are mounted on optical tables and use bulky, not portable lasers which so far has limited the application potential of PU in geology, biology and medicine.

Therefore, it is very attractive to develop a portable PU setup with the elements integrated on a single chip which would facilitate its flexible application to an object and perform PU in situ during a short exposure time. The main element of such a prospective device is the ultrafast pulsed laser providing pulses with a duration ≤1 ps and sufficiently high power, but most importantly this laser should have a compact size to be flexibly transportable. Only recently such miniaturized mode-locked lasers based on semiconductor strip devices reliably operational with parameters that are attractive for PU experiments were designed [15-17]. Therefore, the task to establish PU pump-probe experiments exploiting these compact lasers can be tackled.

In the present paper we describe pump-probe PU experiments with mode-locked semiconductor strip lasers. As specimen for the PU studies we use an Al metal film which is widely used as photo-elastic transducer in PU experiments [18]. The measured pump-probe signals have similar signal to noise ratios as the signals obtained with a conventional PU setup based on commercial mode-locked titanium-sapphire laser. Finally, we discuss various possibilities for the development of a single chip based PU device.

2. Mode-Locked Laser Device

In the experiments we used a laser fabricated by a MOCVD-grown AlGaAs/GaAs epilayer structure with an InGaAs-based active quantum-well (QW) region consisting of two 4.4 nm-thick $In_{0.18}Ga_{0.82}As$ QWs separated by a 9 nm-thick $Al_{0.2}Ga_{0.8}As$ barrier. The parameters of the layers and the layer sequence are presented in Table 1. The active region, designed to emit at a wavelength of around 930 nm, is sandwiched between two 120 nm graded-index AlGaAs layers and two $Al_{0.32}Ga_{0.68}As$ claddings to provide electron and optical confinement. The lower AlGaAs cladding contains a graded far-field reduction layer that enlarges the mode size in the vertical direction, thus reducing both the vertical beam divergence and the optical power density [15].

**Table 1.** Epitaxial layer structure of the AlGaAs-based wafer.

| Layer Type | Materials | THKNS um | Dopant cm-3 |
|---|---|---|---|
| Top Cladding | | | Zinc |
| CAP layer | GaAs | 0.1 | $10^{19} \div 10^{20}$ |
| Matching | $Al_{0.05}Ga_{0.95}As$ | 0.12 | $10^{18} \div 5*10^{18}$ |
| p-cladding | $Al_{0.32}Ga_{0.68}As$ | 1.7 | $10^{18}$ |
| p-cladding | $Al_{0.32}Ga_{0.68}As$ | 0.2 | $5*10^{17} \div 10^{18}$ |
| Active | | | none |
| Graded Index | $Al_{0.2}Ga_{0.8}As$- $Al_{0.32}Ga_{0.68}As$ | 0.12 | - |
| QW | $In_{0.12}Ga_{0.88}As$ | 0.0044 | - |
| Barrier | $Al_{0.2}Ga_{0.8}As$ | 0.009 | - |
| QW | $In_{0.12}Ga_{0.88}As$ | 0.0044 | - |
| Graded Index | $Al_{0.27}Ga_{0.73}As$-$Al_{0.2}Ga_{0.8}As$ | 0.07 | - |
| Bottom | | | Silicon |
| n-cladding | $Al_{0.3}Ga_{0.7}As$- $Al_{0.27}Ga_{0.73}As$ | 0.03 | $10^{17}$ |
| n-cladding | $Al_{0.32}Ga_{0.68}As$- $Al_{0.27}Ga_{0.73}As$ | 0.02 | $5*10^{17} \div 10^{17}$ |
| n-cladding | $Al_{0.32}Ga_{0.68}As$ | 0.75 | $5*10^{17}$ |
| F.F.R layer | $Al_{0.29}Ga_{0.71}As$- $Al_{0.32}Ga_{0.68}As$ | 0.35 | $5*10^{17}$ |
| F.F.R layer | $Al_{0.32}Ga_{0.68}As$- $Al_{0.29}Ga_{0.71}As$ | 0.35 | $5*10^{17}$ |
| n-cladding | $Al_{0.32}Ga_{0.68}As$ | 1.6 | $7*10^{17}$ |
| Matching | $Al_{0.05}Ga_{0.95}As$- $Al_{0.32}Ga_{0.68}As$ | 0.2 | $10^{18} \div 7*10^{17}$ |
| Substrate | GaAs | 645 | Silicon |

The cross section of the laser is shown in Fig.1(a). The optical waveguide is fabricated by optical lithography and reactive ion etching with a $SiCl_4$ gas chemistry [19]. The waveguide has a width of 2.5 μm and an etching depth of approximately 2.1 μm to ensure single spatial mode operation. The total cavity length of 2.6 mm provides a cavity round trip time of 62 ps, corresponding to a repetition rate of 16 GHz under mode-locking conditions. The typical power-current voltage characteristics for such type of lasers is shown in Fig. 1(b). The passive mode-locking is achieved by introducing an intensity dependent loss element into the laser cavity, i.e., a saturable absorber. The absorber consists of an additional reversely biased section along the laser cavity [16]. Because of the saturation arising from the band filling effect, pulses as short as 1 ps can be generated under appropriate biasing conditions [17].

In the PU experiments we used a bar which consists of 20 lasers with identical design and similar characteristics. An optical image of a fragment of the bar is shown in Fig. 1 (c). The laser emission spectra for different absorber voltages measured at a bias current of 120 mA are shown in Fig.1 (d). It is seen that the emission shifts to shorter wavelengths and the spectrum broadens with increasing voltage. In the PU experiments we used 1.5 V on the absorber so that the laser emits at 928 nm.

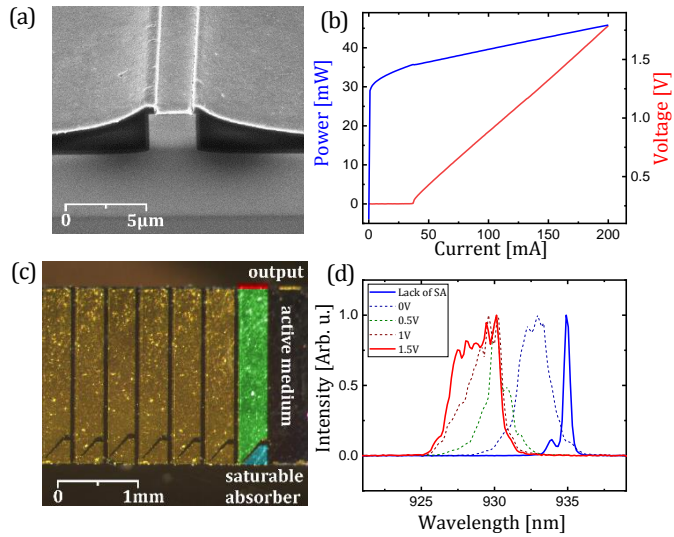

**Fig. 1.** Mode locked semiconductor laser: (a) cross section of the laser; (b) current voltage and output power characteristics of the laser with zero bias on the absorber; (c) optical image of the laser bars on the chip; (d) emission spectra of the laser for various biases applied to the absorber.

### 3. Experiment

The schematic of the PU pump probe setup is shown in Fig.2 (a). The laser beam is collimated by the aspherical lens with focus distance of 3 mm. The beam is split into two paths for pump and probe. The pump passes the chopper for frequency modulation and the probe is sent to the shaker in order to achieve a variable time delay between the pulses in a range of up to 50 ps. Subsequently both beams go through the x15 reflective microscope objective and are focused to a spot with a diameter of 5 μm on the Al film as shown in Fig. 2(a). The average pump power used is 15 mW, which corresponds to ~4 μJ/cm2 excitation density on the Al film. The beams are reflected from the sample, the pump is blocked and the probe beam is sent onto the silicon photodiode. For precise overlap on the sample, an optimized visualization scheme with a microscope set-up was developed. The mirror controls allow one to adjust the pump and probe beam paths and to control precisely the position of the two spots on the sample.

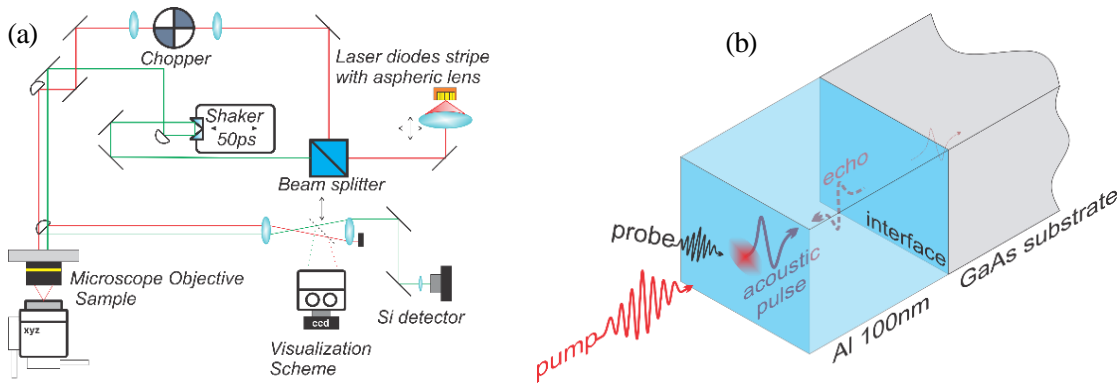

**Fig. 2.** Experimental setup for picosecond ultrasonic experiments: (a) experimental schematic; (b) scheme demonstrating the generation and detection of the echo strain pulse in the Al film.

Figure 3(a) shows the measured PU signal recorded with the semiconductor laser described above. The pump pulse excites the Al film at $t=0$. The strong response observed in the time interval between $t=0$ and 15 ps contains contributions from hot electrons which makes it difficult to extract the contribution from the generated strain. The negative pulse with a maximum at $t=36$ ps corresponds to the first echo of the strain pulse reflected at the Al/GaAs interface as shown in the inset of Fig. 2(b). Indeed, the arrival time of the maximum is expected at $t_0=2d/s =35$ ps, where $d=112$ nm is the nominal thickness of the Al film and $s=6420$ m/s is the velocity of longitudinal sound in Al. The slight difference of 1ps between the expected and measured values of the echo arrival time is due to a slight difference between the actual and nominal thicknesses of the Al film. The reflection $R$ of the acoustic wave at the Al/GaAs interface depends on the acoustic impedances $Z=\rho s$ of the materials, where $\rho$ is the density:

$$R = \frac{Z_{GaAs}-Z_{Al}}{Z_{GaAs}+Z_{Al}} \quad (1)$$

Substituting the values for $Z$ into Eq. (1) we get $R=0.2$ and the reflected strain pulse does not change the phase at the Al/GaAs interface. If there was no contribution from the hot electrons at time around $t=0$, we could expect a negative strain pulse with an amplitude 5 times larger than the amplitude of the echo signal. At $t=0$ the pump-probe signal is positive which means that the contribution from the hot electrons is several times larger than the expected negative contribution from the photoelastic effect.

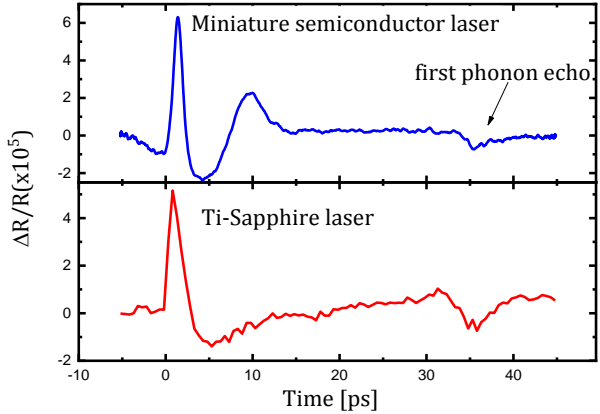

**Fig. 3.** Picosecond ultrasonic signals: (a) measured with the mode-locked semiconductor laser with repetition rate 16 GHz; (b) measured with 80 MHz commercial Ti-sapphire laser.

The PU signal in Fig. 3(a) measured with the miniaturized semiconductor laser has a shape similar to the PU signals obtained in earlier experiments with Al films [18]. To be sure that our results obtained with the semiconductor laser are basically identical to those obtained with a conventional femtosecond laser, we performed PU experiments on the same Al film with a Ti-sapphire laser with 82 MHz repetition rate at 900 nm wavelength and 150 fs pulse duration. However, the excitation energy density in the pump pulse in this case is an order of magnitude higher, resulting in a stronger temperature rise of the Al film. The measured pump-probe trace is shown in Fig. 3(b). We see that the echo signal has a similar temporal shape and appears exactly at the same time as in the experiments with the miniaturized semiconductor laser. The measurement times in both experiments were the same, around 5 minutes, and the signal to noise ratios are almost the same. There is a difference in the temporal shapes around $t=0$ which is likely due to the 200-fold difference in repetition rates and the incomplete recovery of the Al film to lattice temperature in the experiment with the semiconductor laser.

**4. Discussion**

Our experiments show that miniaturized mode-locked semiconductor lasers are very well suitable for PU experiments. The typical property of these lasers is the high repetition rate which is given by the length of the laser cavity and typically is in the range of tens of GHz. The high repetition rate has the advantage of getting the required signal to noise ratio in a shorter time in comparison with conventional mode-locked lasers with repetition rates ~100 MHz, using the same pump fluence in a single pulse. However, the disadvantage of the high-repetition rate is that the system potentially has no time to recover to the initial state. As we see in our experiments this does not have a significant effect on the echo signal, but could become important when studying PU signals that last longer than hundreds of picoseconds. The challenge of getting short laser pulses in miniaturized lasers with durations below a few ps may not be met with every laser and requires careful selection, but our experiments show that a temporal resolution of $\Delta t \approx 1$ ps can be achieved. The duration of the optical pulses which limits the temporal resolution of the PU experiments limits also the highest frequency of the detected acoustic wave. With durations of ~1 ps for the pump and probe pulses, acoustic frequencies of several hundreds of GHz are achievable. PU experiments in this frequency range are high on demand in nondestructive material characterization and imaging of biological objects, including live cells.

The miniaturized mode-locked semiconductor lasers for PU experiments may have even higher average power than the one used so far: While this device relies on a standard two-segment ridge waveguide utilizing quantum well gain, great promise is held by implementing quantum dots as active gain material into the laser. Quantum dots offer several advantages for mode-locked operation of lasers such as a broad gain spectrum, a low linewidth enhancement factor, and ultrafast dynamics with low saturation energies. A second crucial step is the implementation of tapered mode-locked lasers, which consist of three segments: a saturable absorber to enable mode-locked operation, a ridge waveguide and a taper with an opening angle of a few degrees. In gain guided tapered lasers, the mode defined by the ridge waveguide is allowed to diffract into the taper without any lateral index discontinuities, under an angle (typically a few degrees) defined by the mode profile of the ridge. The width of the taper at the facet is an order of magnitude larger than that of a standard ridge waveguide, leading to a reduction of the power density and therefore the facet load. In comparison to simple broad area lasers, which can provide only a moderate beam quality, tapered lasers offer a very reasonable beam quality with beam parameter products ($M^2$) below 2. This allows tight focusing of the output beam from a tapered laser. In addition to the advantages for mode locked operation, quantum dot active layers have also benefits for tapered lasers, in particular the small linewidth enhancement factor, which reduces beam filamentation in the taper and therefore increases the beam quality of the output.

In order to test this, we utilized an MBE-grown laser structure based on AlGaInAs quantum dot laser material, emitting at ~920 nm. The QDs were embedded in a 250 nm thick AlGaAs graded index heterostructure (GRINSCH), and a symmetric AlGaAs cladding (see Table 2). The waveguide structures were defined by electron-beam lithography and reactive ion etching, yielding devices with an overall length of 2.250 mm and an absorber length ranging from 50 to 400 µm. In order to observe the transition from normal to mode-locked operation, the output of the laser was analyzed using a fast fiber-coupled optical receiver and a microwave spectrum analyzer. A significant narrowing of the RF-spectra was observed when reverse bias was applied to the absorber segment, indicating the onset of mode-locking. The peak of the RF frequency (see Fig. 4) is located at 18.05 GHz, corresponding to a cavity round trip time of 55.4 ps.

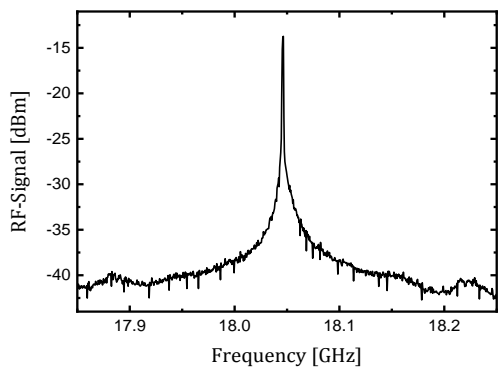

**Fig. 4.** Microwave spectrum of the mode-locked quantum dot laser.

**Table 2.** Epitaxial layer structure of the QD based wafer.

| Layer Type | Materials | THKNS µm | Dopant |
|---|---|---|---|
| Top Cladding | | | Beryllium |
| CAP layer | GaAs | 0.25 | |
| p-cladding | $Al_{0.40}Ga_{0.60}As$ | 1.5 | |
| Active Medium | | | none |
| GRINSH | GaAs/AlGaAs SSL, $Al_{0.18}Ga_{0.82}As$ | 0.25 | - |
| Barrier | GaAs | 0.003 | - |
| SK QD 4.9ML | $Al_{0.15}Ga_{0.23}In_{0.62}As$ | - | - |
| Barrier | GaAs | 0.003 | - |
| GRINSH | GaAs/AlGaAs SSL, $Al_{0.18}Ga_{0.82}As$ | 0.25 | - |
| Bottom Cladding | | | Silicon |
| n-cladding | $Al_{0.40}Ga_{0.60}As$ | 1.5 | |
| Substrate | GaAs | - | Silicon |

Implementing a mode-locked miniaturized laser is the first step in creating a portable PU setup. Another quite massive element in standard PU setups is the delay line. The best solution there will be using Asynchronous Optical Sampling (ASOPS) [20], which requires two mode-locked semiconductor lasers.

This will pave the way for realizing portable sub-THz acoustic equipment for in-situ picosecond ultrasonic studies.

In conclusion, we have demonstrated picosecond ultrasonic setup on the basis of miniaturized mode-locked semiconductor laser. The measurement of a picosecond acoustic echo in Al film shows that the ultrasonic signal has the same shape and signal to noise ratio as in conventional setups with bulky mode-locked lasers. The experiments show the prospective of designing and developing portable picosecond ultrasonic setups integrated on a chip for utilization in medicine, biology and geology.

**CRediT authorship contribution statement**

**Michal Kobecki, Alexey V. Scherbakov, Thomas Czerniuk**: Picosecond ultrasonic setup and measurements, preparing manuscript. **Giuseppe Tandoi, Eugenio Di Gaetano, Marc Sorel**: Fabrication of MOCVD lasers; editing manuscript. **Christian Schneider, Martin Kamp, Sven Höfling**: Fabrication of MBE lasers; editing manuscript. **Andrey V. Akimov**, Conceptualization, first draft of the manuscript. **Manfred Bayer**: Supervision, conceptualization, editing the manuscript

**Acknowledgement**

The work was supported by the Bundesministerium für Bildung und Forschung through the project "Nano-Akusto-Mechanik mit integriertem Laser (NAMIL)" (FKZ 13N14071) and the Deutsche Forschungsgemeinschaft in the frame of TRR 142 (project A6). We are thankful to Dirk Schemionek for depositing the Al films for the picosecond acoustic experiments.